\documentclass[useAMS,usenatbib]{mn2e}

\usepackage[dvips]{graphicx}
\usepackage{amssymb}
\usepackage{amsmath}
\usepackage{amstext}
\usepackage{multirow} 
\usepackage{lscape}

\newcommand{\elecd}{$n_{\rm e}$}
\newcommand{\elect}{$T_{\rm e}$}
\newcommand{\tf}{$t^{2}$}
\newcommand{\hb}{H$\beta$}
\newcommand{\ha}{H$\alpha$}

\newcommand{\foiii}{[O\thinspace{\sc iii}]}

\newcommand{\foi}{[O\thinspace{\sc i}]}
\newcommand{\foii}{[O\thinspace{\sc ii}]}
\newcommand{\fsii}{[S\thinspace{\sc ii}]}
\newcommand{\fsiii}{[S\thinspace{\sc iii}]}
\newcommand{\fni}{[N\thinspace{\sc i}]}
\newcommand{\fnii}{[N\thinspace{\sc ii}]}
\newcommand{\fariv}{[Ar\thinspace{\sc iv}]}
\newcommand{\fcliii}{[Cl\thinspace{\sc iii}]}
\newcommand{\fclii}{[Cl\thinspace{\sc ii}]}
\newcommand{\fcliv}{[Cl\thinspace{\sc iv}]}
\newcommand{\fneiii}{[Ne\thinspace{\sc iii}]}
\newcommand{\ffeii}{[Fe\thinspace{\sc ii}]}
\newcommand{\ffeiii}{[Fe\thinspace{\sc iii}]}
\newcommand{\ffeiv}{[Fe\thinspace{\sc iv}]}

\newcommand{\nitroi}{N\thinspace{\sc i}}
\newcommand{\nii}{N\thinspace{\sc ii}}

\newcommand{\silii}{Si\thinspace{\sc ii}}
\newcommand{\oi}{O\thinspace{\sc i}}
\newcommand{\oii}{O\thinspace{\sc ii}}

\newcommand{\cii}{C\thinspace{\sc ii}}

\newcommand{\niqii}{Ni\thinspace{\sc ii}}
\newcommand{\fciii}{C\thinspace{\sc iii}]}

\newcommand{\fariii}{[Ar\thinspace{\sc iii}]}

\newcommand{\feiv}{Fe\thinspace{\sc iv}}

\newcommand{\hi}{H\,{\sc i}}
\newcommand{\hii}{H\thinspace{\sc ii}}

\newcommand{\hei}{He\thinspace{\sc i}}

\newcommand{\ts}{\emph{$t^2$}}

\newcommand{\ngc}{NGC~2579}
\newcommand{\adfo}{${\rm ADF(O^{2+})}$}

\newcommand\ion[2]{${\rm #1^{#2}}$}           

\newcommand{\cmc}{{\rm cm$^{-3}$}}

\title[\ngc\ and the Galactic abundance gradients]{\ngc\ and the carbon and oxygen abundance gradients beyond the solar circle%
			       \thanks{Based on observations collected at the European Southern 
			       Observatory, Chile, proposal number ESO 382.C-0195(A).}}
\author[C. Esteban et al.]
       {C. Esteban$^{1,2}$\thanks{E-mail: cel@iac.es}, 
        L. Carigi$^3$, M. V. F. Copetti$^{4,5}$, J. Garc{\'{\i}}a-Rojas$^{1,2}$, A. Mesa-Delgado$^6$,    
	\newauthor 
	H. O. Casta\~neda$^7$, D. P{\'e}quignot$^5$ \\
	$^1$Instituto de Astrof\'\i sica de Canarias, E-38200 La Laguna, Tenerife, Spain\\
        $^2$Departamento de Astrof\'\i sica, Universidad de La Laguna, E-38206, La Laguna, Tenerife, Spain\\
        $^3$Instituto de Astronom\'\i a, Universidad Nacional Aut\'onoma de M\'exico, Apdo. Postal 70-264, 04510 M\'exico D.F., Mexico\\
        $^4$Laborat\'orio de An\'alise Num\'erica e Astrof\'\i sica, Departamento de Matem\'atica, Universidade Federal de Santa Maria,\\    
           \ 97119-900 Santa Maria, RS, Brazil\\
        $^5$LUTH, Observatoire de Paris-Meudon, 92190 Meudon, France\\
        $^6$Departamento de Astronom\'ia y Astrof\'isica, Facultad de F\'isica, Pontificia Universidad Cat\'olica de 
        		Chile, Av.~Vicu\~na Mackenna 4860,\\ 782-0436 Macul, Santiago, Chile\\
        $^7$Departamento de F\'\i sica, Escuela Superior de F\'\i sica y Matem\'atica, Instituto Polit\'ecnico Nacional, M\'exico D.F., Mexico\\}

\begin{document}

\date{Accepted 2013 April 25. Received 2013 April 24; in original form 2013 March 13}
\pagerange{\pageref{firstpage}--\pageref{lastpage}} \pubyear{2010}

\maketitle
\label{firstpage}

\begin{abstract}
 We present deep echelle spectrophotometry of the Galactic {\hii} region {\ngc}. The data have been taken with the Very Large Telescope Ultraviolet-Visual Echelle Spectrograph in the 
 3550--10400 \AA\ range. This object, which has been largely neglected, shows however a rather high surface brightness, a high ionization degree and is located at a galactocentric 
 distance of 12.4 $\pm$ 0.7 kpc. Therefore, {\ngc} is an excellent probe for studying the behaviour of the gas phase radial abundance gradients in the outer disc of the Milky Way. We derive the 
 physical conditions of the nebula using several emission line-intensity ratios as well as the abundances of several ionic species from the intensity of collisionally excited lines. We 
 also determine the ionic abundances of C$^{2+}$, O$^+$ and O$^{2+}$ -- and therefore the total O abundance -- from faint pure recombination lines. The results for {\ngc} permit to extend our 
 previous determinations of the C, O and C/O gas phase radial 
 gradients of the inner Galactic disc \citep{estebanetal05} to larger galactocentric distances.  We find that the chemical composition of {\ngc} is consistent with flatten gradients at 
 its galactocentric distance. In addition, we have built a tailored chemical evolution model that reproduces the observed radial abundance gradients of O, C and N and other observational constraints. We find that a levelling out of the star formation efficiency about and beyond the isophotal radius can explain the flattening of chemical gradients observed in the outer Galactic disc.
 
\end{abstract}

\begin{keywords}
 Galaxy: abundances -- Galaxy: evolution -- ISM: abundances -- {\hii} regions -- ISM: individual: \ngc
\end{keywords}

\section{Introduction} \label{intro}

High-resolution spectroscopy with large aperture telescopes has permitted to 
measure very faint pure recombination lines (hereafter RLs) of heavy-element ions in Galactic and 
extragalactic {\hii} regions \citep[see][]{garciarojasesteban07, apeimbert03, lopezsanchezetal07, estebanetal09}. The brightest of these RLs 
are {\cii} 4267 \AA\ and those of multiplet 1 of {\oii} about 4650 \AA\ that can be used to determine the 
ionised gas phase \ion{C}{2+} and \ion{O}{2+} abundances, respectively. In {\hii} regions, a systematic result 
is that abundances obtained 
with RLs are {\it always} between 1.5 to 3 times larger (the so-called abundance discrepancy factor, ADF) than 
those determined using the much brighter collisionally excited lines (hereafter CELs) of the same ions,  
{\fciii} 1909 \AA\ in the UV and those of {\foiii} in the optical \citep[e.g.][]{garciarojasesteban07,estebanetal09}. 
The origin of the abundance discrepancy (hereafter AD) is still an unsolved 
problem. \citet{torrespeimbertetal80} suggested that the presence of fluctuations in the spatial 
distribution of electron temperature can produce 
such discrepancy but others as \citet{tsamispequignot05} and \citet{stasinskaetal07} consider that the AD may be produced by 
small spatial scale chemical inhomogeneities in the interstellar medium. \citet{tsamisetal10} 
have proposed that the presence of small and dense partially-ionised clumps along the line of sight are affecting 
the CEL-derived abundances, producing unrealistic lower values with respect to the RL-based ones, and a similar conclusion 
is found in the results by \citet{mesadelgadoetal12}. More recently, a new hypothesis  based on the presence of a $\kappa$-distribution 
in the energy of the free electrons has been proposed to explain the AD \citep{nichollsetal12}.
 
From deep echelle spectra of a sample of Galactic {\hii} regions, {\citet{estebanetal05} determined the carbon and oxygen 
gradients of the Galactic disc at galactocentric distances between 6.3 and 10.4 kpc based on the intensity of 
RLs. \cite{carigietal05} built chemical evolution models in order to reproduce those abundance gradients 
as well as other observational constraints. They found 
the important result that only models including carbon yields that increase with metallicity for massive stars and 
decrease with metallicity for low and intermediate mass stars can successfully reproduce simultaneously the gradients 
of both elements. Other result of that paper is that the fraction of carbon produced by massive stars 
with respect to that produced by low and intermediate mass stars strongly depends on the age of the Galactic disc, 
as well as on the galactocentric distance. Therefore, it is clear that similar C and O abundance data of {\hii} 
regions at larger galactocentric distances are necessary to refine and extend chemical evolution models and explore possible changes across the Galactic disc. 

There are some evidences that gas phase abundance gradients may flatten out at the outer 
parts of the Galactic disc \citep{fichsilkey91,vilchezesteban96,macielquireza99,costaetal04,macieletal06}. 
However, this result is not supported by other studies \citep{deharvengetal00,rudolphetal06,henryetal10,balseretal11}. 
Chemical evolution models of the Galactic 
disc also obtain different solutions for the behaviour of the abundance gradients at large galactocentric 
distances. 
For example, \citet{fuetal09} find a steepening of the gradients as radial distances increase and  
\citet{marcon-uchidaetal10} can predict a flattening or a steeping depending on the assumed behaviour of the star 
formation efficiency across the Galactic disc. Moreover, chemodynamical models of \citet{samlandetal97} also predict the 
development of a plateau in the outer regions of the Galaxy. 
Therefore, the finding of indisputable observational evidence of the 
presence or not of such flattening is essential to increase our knowledge on important ingredients of chemical evolution models such as the timescale for disc formation and the density threshold for star formation. 

{\ngc} is a high surface brightness Galactic {\hii} region that has been largely neglected because its 
misclassification as either planetary nebula \citep{ackeretal92} or reflection nebula \citep{vanderberghherbst75} 
and even its confusion with other nearby objects \citep{archinalhynes03}. \citet{copettietal07} have brought into 
attention the interest of {\ngc} presenting the first comprehensive observational study of the nebular properties and 
stellar content of this object. These authors derive consistent photometric and kinematic distances for {\ngc} 
indicating that it is located at a galactocentric distance of 12.4 $\pm$ 0.7 kpc (assuming the Sun at 8 kpc from the 
Galactic Centre). Therefore, {\ngc} is probably the brightest 
{\hii} region located  at that distant part of the Galaxy, well outside the solar circle. Another remarkable  
characteristic of this nebula is its relatively high ionization degree, which facilitates the determination 
of chemical abundances and the determination of C$^{2+}$ and O$^{2+}$ abundances 
from RLs. Undoubtedly, {\ngc} is an excellent probe for the exploration of 
the behaviour of the ionised gas phase radial abundance gradients in the external parts of the Galactic disc. 

In \S\ref{observations} we describe the observations and the data reduction procedure. 
In \S\ref{lines} we describe the emission line measurements and identifications as well as the reddening correction. 
In \S\ref{results} we present the physical conditions and ionic and total abundances determined for {\ngc}. 
In \S\ref{discussion} we describe the ingredients of our chemical evolution model and discuss the results in the light of the radial Galactic abundance gradients. 
Finally, in \S\ref{conclusions} we summarize our main conclusions. 
 
  \begin{figure}
   \centering
   \includegraphics[scale=0.45]{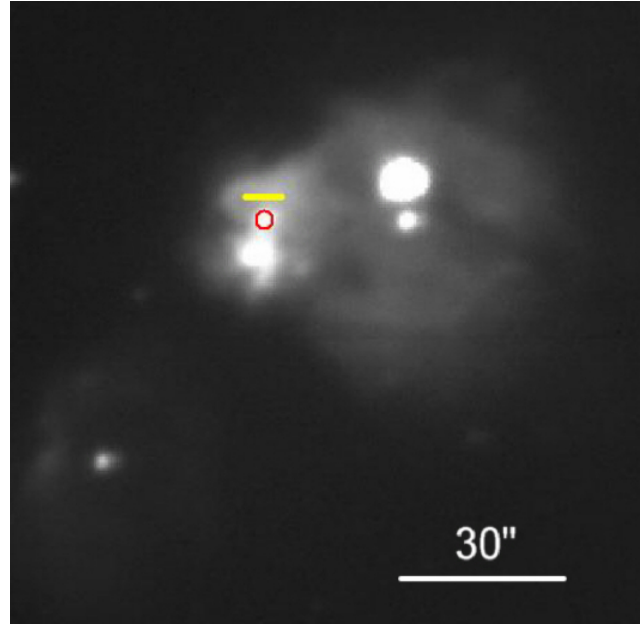} 
   \caption{Combination of {\ha} and continuum image of {\ngc}. The position of
the reference star DENIS J082054.8-361258 is indicated with a small circle. The grey (yellow) horizontal line
centered at 5 arcsec to the north of the reference star indicates the area from which we extracted our 
one-dimensional spectra. North is up and east to the left.}
   \label{f1}
  \end{figure}

\section{Observations and Data Reduction} \label{observations}
 {\ngc} was observed in service time on 2009 January 23 at Cerro Paranal Observatory (Chile), using the UT2 (Kueyen) 
 of the Very Large Telescope (VLT) with the Ultraviolet Visual Echelle Spectrograph 
 \citep[UVES,][]{dodoricoetal00}. The standard settings of UVES were used covering the spectral 
 range from 3570 to 10400 \AA. Some small spectral intervals could not be observed. These are: 
 5734-5833 and 8425-8768 due to the physical separation between the CCDs of the detector 
 system of the red arm; and several much smaller wavelength ranges between 8800 to 10400 \AA\ 
 because the last orders of the spectrum do not fit completely within the size of the CCD. 
 
Three consecutive exposures of 400 
 seconds -- for the 3570-3840 and 4820-6740 \AA\ ranges -- and of 1800 seconds -- for the 3770-4940 and 
 6735-10400 \AA\ ranges -- each were added to obtain the final spectra. In addition, exposures of 30 and 
 60 seconds were taken to obtain non-saturated flux measurements for the brightest 
 emission lines. The full width at half-maximum (FWHM) at a given 
 $\lambda$ was $\Delta\lambda\approx\lambda/8000$. The slit was oriented east-west and the
 atmospheric dispersion corrector (ADC) was used to keep 
 the same observed region within the slit regardless of the air mass value. The slit width was set 
 to 3$\arcsec$ as a compromise between the spectral resolution needed and the desired signal-to-noise 
 ratio of the spectra. The slit length was fixed to 10$\arcsec$. The one-dimensional spectra we finally analysed were 
 extracted for an area of 3$\arcsec\times$7\farcs4. This area covers the brightest part of the 
 nebula (see Figure~\ref{f1}), located 5$\arcsec$ to the north of the star DENIS J082054.8-361258, 
 a member of the stellar cluster that ionises the nebula. 

 The spectra were reduced using the {\sc iraf}\footnote{{\sc iraf} is distributed by NOAO, which 
 is operated by AURA, under cooperative agreement with NSF.} echelle reduction package, following 
 the standard procedure of bias subtraction, aperture extraction, flatfielding, wavelength 
 calibration and flux calibration. The standard star LTT 3218 \citep{hamuyetal92,%
 hamuyetal94} was observed to perform the flux calibration. 
  \begin{figure}
   \centering
   \includegraphics[scale=0.45]{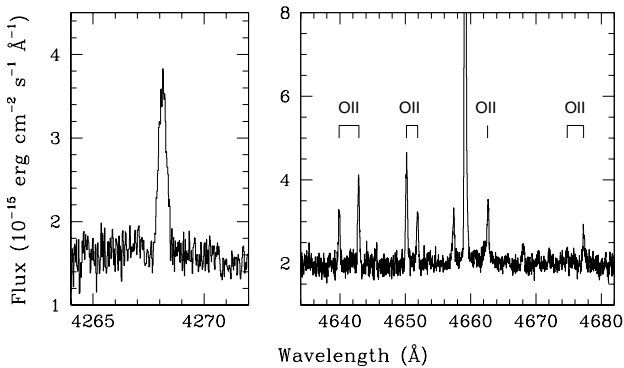} 
   \caption{Sections of the UVES spectrum on {\ngc} -- uncorrected for reddening -- showing the faint recombination 
            lines of {\cii} 4267 \AA\ (left) and multiplet 1 of {\oii} about 4650 \AA\ (right).}
   \label{f2}
  \end{figure}    	 

\section{Line intensities} \label{lines}
 Line fluxes were measured by integrating all the flux in the line between two given 
limits and over a local continuum estimated by eye. In the case of line blending, a  
double Gaussian profile fit procedure was applied to measure the 
individual line intensities. All these measurements were made with the {\sc splot} routine of {\sc iraf}.

 All line fluxes of a given spectrum have been normalized to a particular bright emission line 
 in each spectral range. For the bluest range (3570-3840 \AA), the reference line was H12 3750 \AA. 
 For the 4820-6740 and 3770-4940 \AA\ ranges, H$\beta$ was used. For the reddest 
spectral interval (6735-10400 \AA), the reference was P20 8392 \AA. In order to produce a final homogeneous 
set of line flux ratios, all of them were rescaled to H$\beta$ considering the theoretical 
H12/H$\beta$ and P20/H$\beta$ line ratios corresponding to the physical conditions of the gas. 

 The spectral ranges present some overlapping at the edges. The final flux of a line in the 
 overlapping regions was the average of the values obtained in both adjacent spectra. A similar procedure 
 was considered in the case of lines present in two consecutive spectral orders. The average of both 
 measurements were considered as the adopted value of the 
 line flux. In all cases, the differences in the line flux measured for the same line in 
 different orders and/or spectral ranges do not show systematic trends and are always within the 
 uncertainties.

 The identification and laboratory wavelengths of the lines were obtained following our previous works 
 on echelle spectroscopy of bright Galactic \hii\ regions \cite[see][and references therein]{garciarojasesteban07}. 
 We also used a preliminar (unpublished) version of the spectral synthesis code for nebulae  X-SSN \citep{pequignotetal12} to identify faint lines. 
 This code was specially useful to identify lines in the reddest part of the spectrum, where confusion with telluric features 
 is problematic. 

 For a given line, the observed wavelength is determined by the centre of the baseline chosen for the 
flux integration procedures or the centroid of the line when a Gaussian fit is used (in the case of 
line blending). The final adopted values of the observed wavelength of a given line are relative 
to the heliocentric reference frame. 

 The reddening coefficient, $c$({\hb}), was determined from the comparison of the observed flux 
 ratio of the brightest Balmer lines -- H$\alpha$, H$\gamma$, H$\delta$ and H$\nu$ lines with respect to H$\beta$ -- and the theoretical line ratios computed 
 by \cite{storeyhummer95} for the physical conditions of the nebula. We have used 
 the reddening function, $f(\lambda)$, normalized to H$\beta$ derived by \cite{cardellietal89} and assuming 
 $R_V$ = 3.1. 
 The c(H$\beta$) we obtain is 2.37$\pm$0.05, which is larger than the values of 
 the reddening coefficient determined by \cite{copettietal07} who use the reddening function 
 by \cite{kaler76} and obtain a range of values from 1.12 to 1.65 depending on the position 
 inside the nebula and the Balmer lines used. 
  
 In Table~\ref{table1}, the final list of line identifications (columns 1--3), observed wavelength 
 corrected for heliocentric velocity (column 4), observed and dereddened flux line ratios 
 with respect to H$\beta$ (columns 5 and 6) and the uncertainty of line ratios -- in percentage -- 
 (last column) are presented. The quoted errors include the uncertainties in line 
 flux measurement and error propagation in the reddening coefficient.
 
 In Figure~\ref{f2}, we show sections of our flux-calibrated echelle spectra showing the recombination 
 lines of {\cii} 4267 \AA\ and multiplet 1 of {\oii} around 4650 \AA. As it can be seen, these faint lines are well separated and 
 show a remarkably high signal-to-noise ratio in our deep echelle spectrum. 
 
  \begin{table*}
 \begin{minipage}{155mm}
 \centering
  \caption{Observed, $F$($\lambda$), and reddening-corrected, $I$($\lambda$), line ratios [$F$(H$\beta$) = 100] for \ngc.}
   \label{table1}	   
  \begin{tabular}{llcrrrr}
  \hline
     $\lambda_0$ & & & $\lambda_{obs}$ & & & Error \\
     (\AA) &  Ion & Multiplet & (\AA) & $F$($\lambda$) & $I$($\lambda$) & (\%)$^a$ \\
  \hline
3587.28	&	{\hei}	&	3	&	3588.08	&	0.039	&	0.180	&	17	\\				
3613.64	&	{\hei}	&	6	&	3614.47	&	0.067	&	0.301	&	12	\\				
3634.25	&	{\hei}	&	28	&	3635.07	&	0.052	&	0.227	&	14	\\				
3662.26	&	{\hi}	&	H30	&	3663.19	&	0.045	&	0.192	&	16	\\				
3663.40	&	{\hi}	&	H29	&	3664.21	&	0.036	&	0.154	&	18	\\				
3664.68	&	{\hi}	&	H28	&	3665.54	&	0.056	&	0.238	&	14	\\				
3666.10	&	{\hi}	&	H27	&	3666.90	&	0.059	&	0.254	&	13	\\				
3667.68	&	{\hi}	&	H26	&	3668.54	&	0.076	&	0.327	&	12	\\				
3669.47	&	{\hi}	&	H25	&	3670.31	&	0.080	&	0.343	&	11	\\				
3671.48	&	{\hi}	&	H24	&	3672.31	&	0.088	&	0.377	&	11	\\				
3673.76	&	{\hi}	&	H23	&	3674.60	&	0.090	&	0.384	&	11	\\				
3676.37	&	{\hi}	&	H22	&	3677.20	&	0.114	&	0.483	&	10	\\				
3679.36	&	{\hi}	&	H21	&	3680.21	&	0.156	&	0.663	&	8	\\				
3682.81	&	{\hi}	&	H20	&	3683.62	&	0.138	&	0.581	&	9	\\				
3686.83	&	{\hi}	&	H19	&	3687.68	&	0.126	&	0.531	&	9	\\				
3691.56	&	{\hi}	&	H18	&	3692.42	&	0.186	&	0.779	&	8	\\				
3697.15	&	{\hi}	&	H17	&	3697.98	&	0.250	&	1.042	&	7	\\				
3703.86	&	{\hi}	&	H16	&	3704.71	&	0.262	&	1.085	&	7	\\				
3705.04	&	{\hei}	&	25	&	3705.87	&	0.121	&	0.501	&	9	\\				
3711.97	&	{\hi}	&	H15	&	3712.81	&	0.325	&	1.338	&	6	\\				
3721.83	&	{\fsiii}	&	2F	&	3722.68	&	0.646	&	2.631	&	6	\\				
3721.94	&	{\hi}	&	H14	&		&		&		&		\\				
3726.03	&	{\foii}	&	1F	&	3726.90	&	16.00	&	64.94	&	5	\\				
3728.82	&	{\foii}	&	1F	&	3729.65	&	12.51	&	50.63	&	5	\\				
3734.37	&	{\hi}	&	H13	&	3735.22	&	0.467	&	1.880	&	6	\\				
3750.15	&	{\hi}	&	H12	&	3751.00	&	0.690	&	2.737	&	6	\\				
3770.63	&	{\hi}	&	H11	&	3771.49	&	0.815	&	3.168	&	5	\\				
3797.90	&	{\hi}	&	H10	&	3798.76	&	1.179	&	4.462	&	5	\\				
3819.61	&	{\hei}	&	22	&	3820.49	&	0.262	&	0.969	&	7	\\				
3833.57	&	{\hei}	&	62	&	3834.47	&	0.017	&	0.062	&	28	\\				
3835.39	&	{\hi}	&	H9	&	3836.26	&	1.710	&	6.229	&	5	\\				
3856.02	&	{\silii}	&	1F	&	3856.90	&	0.026	&	0.093	&	22	\\				
3856.13	&	{\oii}	&	12	&		&		&		&		\\				
3862.59	&	{\silii}	&	1	&	3863.49	&	0.022	&	0.079	&	24	\\				
3867.49	&	{\hei}	&	20	&	3868.34	&	0.042	&	0.150	&	16	\\				
3868.75	&	{\fneiii} 	&	1F	&	3869.64	&	3.409	&	11.99	&	5	\\				
3871.82	&	{\hei}	&	60	&	3872.68	&	0.018	&	0.062	&	27	\\				
3888.65 &       {\hei}	&       2	&	3889.85	&	3.911	&	13.46	&	5	\\
3889.05	&	{\hi}	&	H8	&		&		&		&		\\				
3918.98	&	{\cii}	&	4	&	3919.82	&	0.008	&	0.028	&	:	\\				
3920.68	&	{\cii}	&	4	&	3921.50	&	0.023	&	0.076	&	23	\\				
3926.53	&	{\hei}	&	58	&	3927.42	&	0.035	&	0.114	&	18	\\				
3964.73	&	{\hei}	&	5	&	3965.63	&	0.289	&	0.914	&	6	\\				
3967.46	&	{\fneiii} 	&	1F	&	3968.37	&	1.131	&	3.567	&	5	\\				
3970.07	&	{\hi}	&	H7	&	3970.97	&	4.912	&	15.45	&	5	\\				
4009.22	&	{\hei}	&	55	&	4010.19	&	0.067	&	0.203	&	12	\\				
4026.21	&	{\hei}	&	18	&	4027.12	&	0.733	&	2.161	&	5	\\				
4068.60	&	{\fsii}	&	1F	&	4069.56	&	0.239	&	0.670	&	7	\\				
4069.62	&	{\oii}	&	10	&	4070.66	&	0.033	&	0.092	&	19	\\				
4072.15	&	{\oii}	&	10	&	4073.05	&	0.022	&	0.063	&	24	\\				
4075.86	&	{\oii}	&	10	&	4076.81	&	0.021	&	0.058	&	25	\\				
4076.35	&	{\fsii}	&	1F	&	4077.32	&	0.075	&	0.209	&	12	\\				
4101.74	&	{\hi}	&	H6	&	4102.67	&	9.500	&	25.60	&	5	\\				
4120.82	&	{\hei}	&	16	&	4121.72	&	0.085	&	0.225	&	11	\\				
4143.76	&	{\hei}	&	53	&	4144.72	&	0.136	&	0.347	&	8	\\				
 4153.3 & {\oii}        &     19     &    4154.29 &      0.012 &      0.030 &         36 \\
4168.97	&	{\hei}	&	52	&	4169.96	&	0.019	&	0.048	&	26	\\				
4267.15	&	{\cii}	&	6	&	4268.14	&	0.071	&	0.157	&	12	\\				
4287.40	&	{\ffeii}	&	7F	&	4288.40	&	0.013	&	0.027	&	34	\\				
4317.14	&	{\oii}	&	2	&	4318.07	&	0.010	&	0.021	&	39	\\							
4326.4	&	{\oi}	&		&	4327.45	&	0.010	&	0.021	&	38	\\	
\end{tabular}
\end{minipage}
\end{table*}
\setcounter{table}{0}
\begin{table*}
 \begin{minipage}{155mm}
 \centering
  \caption{continued}
  \begin{tabular}{llcrrrr}
  \hline
     $\lambda_0$ & & & $\lambda_{obs}$ & & & Error \\
     (\AA) &  Ion & Multiplet & (\AA) & $F$($\lambda$) & $I$($\lambda$) & (\%)$^a$ \\
  \hline
4340.47	&	{\hi}	&	H$\gamma$	&	4341.45	&	23.67	&	47.27	&	4	\\				
4345.56	&	{\oii}	&	2	&	4346.57	&	0.019	&	0.038	&	26	\\				
4349.43	&	{\oii}	&	2	&	4350.41	&	0.020	&	0.039	&	25	\\				
4361.60	&	{\nii}	&		&	4362.55	&	0.012	&	0.023	&	35	\\				
4363.21	&	{\foiii}	&	2F	&	4364.19	&	0.968	&	1.877	&	5	\\				
4366.89	&	{\oii}	&	2	&	4367.90	&	0.021	&	0.041	&	24	\\				
4368.19	&	{\oi}	&	5	&	4369.32	&	0.019	&	0.037	&	26	\\				
4368.25	&	{\oi}	&	5	&		&		&		&		\\				
4387.93	&	{\hei}	&	51	&	4388.93	&	0.306	&	0.574	&	6	\\				
4437.55	&	{\hei}	&	50	&	4438.56	&	0.043	&	0.076	&	16	\\				
4471.48	&	{\hei}	&	14	&	4472.52	&	2.727	&	4.592	&	4	\\				
4607.13	&	{\ffeiii}	&	3F	&	4608.13	&	0.023	&	0.032	&	23	\\				
4609.44	&	{\oii}	&	93	&	4610.42	&	0.009	&	0.012	&	:	\\				
4630.54	&	{\nii}	&	5	&	4631.59	&	0.026	&	0.035	&	22	\\				
4638.86	&	{\oii}	&	1	&	4639.90	&	0.037	&	0.051	&	17	\\				
4641.81	&	{\oii}	&	1	&	4642.86	&	0.054	&	0.073	&	14	\\				
4649.13	&	{\oii}	&	1	&	4650.22	&	0.071	&	0.094	&	12	\\				
4650.84	&	{\oii}	&	1	&	4651.87	&	0.034	&	0.045	&	18	\\				
4658.10	&	{\ffeiii}	&	3F	&	4659.18	&	0.417	&	0.547	&	5	\\				
4661.63	&	{\oii}	&	1	&	4662.64	&	0.044	&	0.057	&	16	\\				
4667.01	&	{\ffeiii}	&	3F	&	4668.04	&	0.017	&	0.022	&	28	\\				
4673.73	&	{\oii}	&	1	&	4674.79	&	0.009	&	0.012	&	:	\\				
4676.24	&	{\oii}	&	1	&	4677.34	&	0.025	&	0.032	&	22	\\				
4701.53	&	{\ffeiii}	&	3F	&	4702.65	&	0.116	&	0.143	&	9	\\				
4711.37	&	{\fariv}	&	1F	&	4712.49	&	0.018	&	0.023	&	27	\\				
4713.14	&	{\hei}	&	12	&	4714.25	&	0.501	&	0.611	&	5	\\				
4733.93	&	{\ffeiii}	&	3F	&	4734.99	&	0.042	&	0.050	&	16	\\				
4740.16	&	{\fariv}	&	1F	&	4741.32	&	0.016	&	0.019	&	29	\\				
4754.83	&	{\ffeiii}	&	3F	&	4755.82	&	0.083	&	0.096	&	11	\\				
4769.60	&	{\ffeiii}	&	3F	&	4770.55	&	0.044	&	0.050	&	15	\\				
4777.88	&	{\ffeiii}	&	3F	&	4778.81	&	0.021	&	0.023	&	25	\\				
4861.33	&	{\hi}	&	{\hb}	&	4862.42	&	100.0	&	100.0	&	4	\\				
4881	&	{\ffeiii}	&	2F	&	4882.15	&	0.180	&	0.176	&	7	\\				
4889.62 & {\ffeii}    &     4F       &    4890.82 &      0.014 &      0.014 &         31 \\
4894.59 & {\oi}        &            &    4895.86 &      0.010 &      0.010 &         38 \\
4921.93	&	{\hei}	&	48	&	4923.05	&	1.202	&	1.109	&	4	\\				
4924.5	&	{\ffeiii}	&	2F	&	4925.67	&	0.024	&	0.022	&	22	\\				
4931.32	&	{\foiii}	&	1F	&	4932.36	&	0.045	&	0.041	&	15	\\				
4958.91	&	{\foiii}	&	1F	&	4960.09	&	139.9	&	122.8	&	4	\\				
4985.9	&	{\ffeiii}	&	2F	&	4986.97	&	0.072	&	0.061	&	12	\\				
4987.2	&	{\ffeiii}	&	2F	&	4988.38	&	0.079	&	0.067	&	11	\\				
5006.84	&	{\foiii}	&	1F	&	5008.03	&	432.8	&	356.2	&	4	\\				
5011.30	&	{\ffeiii}	&	1F	&	5012.46	&	0.063	&	0.052	&	12	\\				
5015.68	&	{\hei}	&	4	&	5016.86	&	2.802	&	2.279	&	4	\\				
5041.03	&	{\silii}	&	5	&	5042.21	&	0.100	&	0.079	&	10	\\				
5055.98	&	{\silii}	&	5	&	5057.21	&	0.172	&	0.133	&	7	\\				
5191.82	&	{\fariii}	&	3F	&	5192.91	&	0.100	&	0.064	&	10	\\				
5197.90	&	{\fni}	&	1F	&	5199.27	&	0.149	&	0.095	&	8	\\				
5200.26	&	{\fni}	&	1F	&	5201.69	&	0.092	&	0.059	&	10	\\				
5261.62 & {\ffeii}    &    19F     &    5263.00 &      0.022 &      0.013 &         :  \\
5270.40	&	{\ffeiii}	&	1F	&	5271.75	&	0.341	&	0.198	&	6	\\				
5412.00	&	{\ffeiii}	&	1F	&	5413.40	&	0.038	&	0.018	&	17	\\				
5517.71	&	{\fcliii}	&	1F	&	5518.98	&	0.961	&	0.415	&	5	\\				
5537.88	&	{\fcliii}	&	1F	&	5539.13	&	0.914	&	0.387	&	5	\\				
5875.64	&	{\hei}	&	11	&	5877.03	&	32.51	&	14.44	&	4	\\				
5941.65 & {\nii}        &     28     &    5943.18 &      0.036 &      0.015 &         17 \\
5978.93	&	{\silii}	&	4	&	5980.31	&	0.141	&	0.057	&	8	\\				
6300.30	&	{\foi}	&	1F	&	6301.91	&	1.164	&	0.359	&	4	\\				
6312.10	&	{\fsiii}	&	3F	&	6313.55	&	5.293	&	1.615	&	4	\\				
6347.11	&	{\silii}	&	4	&	6348.58	&	0.505	&	0.150	&	5	\\ 
6363.78	&	{\foi}	&	1F	&	6365.40	&	0.444	&	0.130	&	5	\\				
6371.36	&	{\silii}	&	2	&	6372.87	&	0.260	&	0.076	&	6	\\				
\end{tabular}
\end{minipage}
\end{table*}
\setcounter{table}{0}
\begin{table*}
 \begin{minipage}{155mm}
 \centering
  \caption{continued}
  \begin{tabular}{llcrrrr}
  \hline
     $\lambda_0$ & & & $\lambda_{obs}$ & & & Error \\
     (\AA) &  Ion & Multiplet & (\AA) & $F$($\lambda$) & $I$($\lambda$) & (\%)$^a$ \\
  \hline
6548.03	&	{\fnii}	&	1F	&	6549.61	&	30.40	&	7.681	&	4	\\				
6562.82	&	{\hi}	&	{\ha}	&	6564.31	&	1161	&	290.0	&	4	\\				
6578.05	&	{\cii}	&	2	&	6579.60	&	0.599	&	0.148	&	5	\\				
6583.41	&	{\fnii}	&	1F	&	6585.02	&	96.78	&	23.79	&	4	\\				
6678.15	&	{\hei}	&	46	&	6679.77	&	15.34	&	3.508	&	4	\\				
6716.47	&	{\fsii}	&	2F	&	6718.03	&	15.70	&	3.487	&	4	\\				
6730.85	&	{\fsii}	&	2F	&	6732.60	&	18.66	&	4.101	&	4	\\				
6734.08 &     {\hei}      &        1/22    &    6735.84 &      0.025 &      0.005 &         22 \\
6734.42 & {\ffeiv}   & & & && \\
6739.8 & {\ffeiv}   &            &    6741.65 &      0.014 &      0.004 &         31 \\
6744.1 &  {\hei}      &        1/22    &    6745.93 &      0.011 &      0.003 &         37 \\
6755.95 & {\hei}      &        1/20    &    6757.60 &      0.013 &      0.004 &         33 \\
6755.99 & {\ffeiv}   & & & && \\
6785.68 & {\hei}       &      1/18      &    6787.67 &      0.020 &      0.005 &         25 \\
6787.21 & {\cii}       &      14      &    6788.84 &      0.010 &      0.003 &         : \\
6791.47 & {\cii}       &      14      &    6793.15 &      0.011 &      0.003 &         37 \\
6804.95 & {\hei}       &      1/17      &    6806.75 &      0.025 &      0.007 &         22 \\
6855.88 & {\hei}       &   1/15   &    6857.73 &      0.023 &      0.006 &         23 \\
6933.99 & {\hei}       &    1/13        &    6935.78 &      0.031 &      0.008 &         19 \\
6989.53 & {\hei}       &     1/12       &    6991.31 &      0.051 &      0.012 &          14 \\
7062.34 & {\hei}       &   1/11   &    7064.10 &      0.085 &      0.019 &          11 \\
7065.28	&	{\hei}	&	10	&	7066.90	&	26.66	&	5.885	&	4	\\				
7135.78	&	{\fariii}	&	1F	&	7137.45	&	65.90	&	13.87	&	4	\\				
7155.16 & {\ffeii}    &    14F     &    7157.11 &      0.088 &      0.018 &          10 \\
7160.56 & {\hei}       &   1/10   &    7162.42 &      0.135 &      0.028 &          8 \\
7231.32 & {\cii}       &     3      &    7233.25 &      0.180 &      0.036 &          7 \\
7236.42	&	{\cii}	&	3	&	7238.15	&	0.508	&	0.100	&	5	\\				
7281.35	&	{\hei}	&	45	&	7283.06	&	3.512	&	0.671	&	4	\\				
7298.04 & {\hei}       &   1/9    &    7299.84 &      0.102 &      0.019 &          10 \\
7318.39	&	{\foii}	&	2F	&	7320.77	&	4.099	&	0.765	&	4	\\				
7319.99	&	{\foii}	&	2F	&	7321.85	&	13.18	&	2.455	&	4	\\				
7329.66	&	{\foii}	&	2F	&	7331.41	&	6.969	&	1.291	&	4	\\				
7330.73	&	{\foii}	&	2F	&	7332.49	&	7.297	&	1.351	&	4	\\				
7377.83 & [{\niqii}]   &     2F     &    7379.90 &      0.123 &      0.022 &          9 \\
7423.64 & {\nitroi}        &     3      &    7425.70 &      0.022 &      0.004 &         24 \\
7442.3 & {\nitroi}        &     3      &    7444.39 &      0.070 &      0.012 &          12 \\
7452.54 & {\ffeii}    &    14F     &    7454.55 &      0.033 &      0.006 &         19 \\
7499.85 & {\hei}       &   1/8    &    7501.77 &      0.238 &      0.040 &          6 \\
7504.94 & {\oii}       &            &    7506.86 &      0.020 &      0.0033 &         25 \\
7513.33 & {\hei}       &            &    7515.25 &      0.017 &      0.0028 &         28 \\
7519.29 &   {\hei}    &            &    7521.52 &      0.036 &      0.006 &         17 \\
7519.49 & {\cii}   & 16.8 & & && \\
7519.93 & {\cii}   & 16.8 & & && \\
7530.54 & {\fcliv}      &    1F        &    7532.51 &      0.027 &      0.0043 &         21 \\
7530.57 & {\cii}   & 16.8 & & && \\
7751.10	&	{\fariii}	&	2F	&	7752.92	&	20.94	&	3.001	&	4	\\				
7774.17 & {\oi}        &     1      &    7776.17 &      0.031 &      0.0044 &         19 \\
7775.39 & {\oi}        &     1      &    7777.32 &      0.017 &      0.0024 &         28 \\
7816.13	&	{\hei}	&	1/7	&	7817.98	&	0.402	&	0.055	&	4	\\				
8015.96 & {\hei}       &      4/20      &    8017.94 &      0.024 &      0.0029 &         23 \\
8035.06 & {\hei}       &      4/19      &    8037.14 &      0.038 &      0.0047 &         17 \\
8045.62 & {\fcliv}      &    1F        &    8047.82 &      0.59 &      0.0072 &         13 \\
8057.55 & {\hei}       &      4/18      &    8059.50 &      0.042 &      0.0051 &          16 \\
8094.11 & {\hei}       &   4/10   &    8096.18 &      0.039 &      0.0046 &         17 \\
8116.44 & {\hei}       &   4/16   &    8118.45 &      0.049 &      0.0058 &          14 \\
8203.86 & {\hei}       &   4/14   &    8205.98 &      0.113 &      0.0126 &          9 \\
8210.72 & {\nitroi}        &     2      &    8213.11 &      0.045 &      0.0051 &          15 \\
8216.34 & {\nitroi}        &     2      &    8218.82 &      0.078 &      0.0087 &          11 \\
8223.14	&	{\nitroi}	&	2	&	8225.21	&	0.206	&	0.023	&	7	\\				
8233.21	&	{\hi}	&	P50	&	8235.13	&	0.090	&	0.0100	&	10	\\				
8234.43	&	{\hi}	&	P49	&	8236.39	&	0.114	&	0.0125	&	9	\\				
8235.74	&	{\hi}	&	P48	&	8237.71	&	0.113	&	0.0124	&	9	\\				
\end{tabular}
\end{minipage}
\end{table*}
\setcounter{table}{0}
\begin{table*}
 \begin{minipage}{155mm}
 \centering
  \caption{continued}
  \begin{tabular}{llcrrrr}
  \hline
     $\lambda_0$ & & & $\lambda_{obs}$ & & & Error \\
     (\AA) &  Ion & Multiplet & (\AA) & $F$($\lambda$) & $I$($\lambda$) & (\%)$^a$ \\
  \hline
8237.13	&	{\hi}	&	P47	&	8239.05	&	0.183	&	0.020	&	7	\\				
8238.61	&	{\hi}	&	P46	&	8240.55	&	0.202	&	0.022	&	7	\\				
8240.19	&	{\hi}	&	P45	&	8242.02	&	0.283	&	0.031	&	6	\\				
8241.88	&	{\hi}	&	P44	&	8244.12	&	0.304	&	0.033	&	6	\\				
8243.69	&	{\hi}	&	P43	&	8245.60	&	0.274	&	0.030	&	6	\\				
8245.64	&	{\hi}	&	P42	&	8247.56	&	0.379	&	0.042	&	5	\\				
8247.73	&	{\hi}	&	P41	&	8249.66	&	0.346	&	0.038	&	5	\\				
8249.2	&	{\hi}	&	P40	&	8251.89	&	0.343	&	0.038	&	5	\\				
8252.4	&	{\hi}	&	P39	&	8254.35	&	0.404	&	0.044	&	5	\\				
8255.02	&	{\hi}	&	P38	&	8257.04	&	0.319	&	0.035	&	6	\\				
8257.85	&	{\hi}	&	P37	&	8259.82	&	0.367	&	0.040	&	5	\\				
8260.93	&	{\hi}	&	P36	&	8262.84	&	0.460	&	0.050	&	5	\\				
8264.28	&	{\hi}	&	P35	&	8266.28	&	0.660	&	0.072	&	5	\\				
8265.7 & {\hei}       &     2/9       &    8267.71 &      0.055 &      0.0059 &          14 \\
8267.94	&	{\hi}	&	P34	&	8269.89	&	0.555	&	0.060	&	5	\\				
8271.93	&	{\hi}	&	P33	&	8273.80	&	0.512	&	0.055	&	5	\\				
8276.31	&	{\hi}	&	P32	&	8278.33	&	1.543	&	0.167	&	4	\\				
8281.12	&	{\hi}	&	P31	&	8283.03	&	0.865	&	0.093	&	5	\\				
8290.56 & {\hei}       &     6/27       &    8292.71 &      0.025 &      0.0027 &         22 \\
8292.31	&	{\hi}	&	P29	&	8294.28	&	0.597	&	0.064	&	5	\\				
8298.83	&	{\hi}	&	P28	&	8300.82	&	0.785	&	0.084	&	5	\\				
8306.11	&	{\hi}	&	P27	&	8308.06	&	1.067	&	0.114	&	4	\\				
8314.26	&	{\hi}	&	P26	&	8316.22	&	1.055	&	0.112	&	4	\\				
8323.42	&	{\hi}	&	P25	&	8325.39	&	1.290	&	0.136	&	4	\\				
8329.86	&	{\hei}	&	6/23	&	8331.76	&	0.038	&	0.0040	&	17	\\	
8333.78	&	{\hi}	&	P24	&	8335.76	&	1.396	&	0.146	&	4	\\				
8345.55	&	{\hi}	&	P23	&	8347.51	&	1.599	&	0.167	&	4	\\				
8359.00	&	{\hi}	&	P22	&	8360.93	&	2.018	&	0.209	&	4	\\				
8361.73 & {\hei}       &   1/6    &    8363.81 &      0.822 &      0.085 &          5 \\
8374.48	&	{\hi}	&	P21	&	8376.43	&	1.994	&	0.205	&	4	\\				
8376.56 & {\hei}       &   6/20   &    8378.62 &      0.066 &      0.0068 &          12 \\
8392.40	&	{\hi}	&	P20	&	8394.36	&	2.325	&	0.237	&	4	\\				
8413.32	&	{\hi}	&	P19	&	8415.24	&	4.827	&	0.487	&	4	\\				
8421.97 & {\hei}       &   6/18   &    8424.04 &      0.118 &      0.012 &          9 \\
8862.79	&	{\hi}	&	P11	&	8862.50	&	11.71	&	1.410	&	4	\\				
8914.77 & {\hei}        &   2/7    &    8917.12 &      0.119 &      0.014 &          9 \\
8930.77 & {\hei}        &  10/11   &    8933.30 &      0.044 &      0.0051 &          15 \\
8996.99	&	{\hei}	&	6/10	&	8998.67	&	0.489	&	0.056	&	5	\\				
9014.91	&	{\hi}	&	P10	&	9017.24	&	7.810	&	0.882	&	4	\\				
9063.32 & {\hei}        &   4/8    &    9065.61 &      0.466 &      0.052 &          5 \\
9068.90	&	{\fsiii}	&	1F	&	9071.05	&	198.7	&	21.94	&	4	\\				
9123.6 & {\fclii}    &     1F     &    9126.03 &      0.131 &      0.014 &          8 \\
9210.28	&	{\hei}	&	6/9	&	9212.52	&	0.635	&	0.066	&	5	\\				
9213.24 & {\hei}       &   7/9    &    9215.60 &      0.239 &      0.025 &          6 \\
9229.01	&	{\hi}	&	P9	&	9231.26	&	19.40	&	2.008	&	4	\\				
9262.67 & {\oi}        &       8     &    9265.07 &      0.021 &      0.0021 &          25 \\
9265.94 & {\oi}        &        8    &    9268.33 &      0.031 &      0.0031 &          19 \\
9463.57 & {\hei}       &   1/5    &    9465.78 &      1.120 &      0.106 &          4 \\
9526.17 & {\hei}       &   6/8    &    9528.41 &      0.465 &      0.043 &          5 \\
9530.60	&	{\fsiii}	&	1F	&	9532.95	&	520.5	&	48.02	&	4	\\				
9545.97	&	{\hi}	&	P8	&	9548.06	&	27.04	&	2.48	&	4	\\				
9625.70 & {\hei}       &      6/8      &    9628.23 &      0.207 &      0.019 &          7 \\
9682.39 & {\hei}       &      9/8      &    9685.01 &      0.041 &      0.0036 &          16 \\
9701.87 & {\ffeiii}   &            &    9704.40 &      0.131 &      0.0113 &          8 \\
9702.65 & {\hei}       &      75      &    9705.29 &      0.172 &      0.015 &          7 \\
9903.46 & {\cii}       &    17.2    &    9906.40 &      0.389 &      0.032 &          6 \\
10027.70	&	{\hei}	&	6/7	&	10030.66	&	2.050	&	0.159	&	4	\\				
10031.2	&	{\hei}	&	7/7	&	10033.94	&	0.657	&	0.051	&	5	\\				
10049.40	&	{\hi}	&	P7	&	10051.76	&	57.70	&	4.450	&	4	\\
  10138.42 & {\hei}       &   10/7   &   10140.91 &      0.247 &      0.019 &          6 \\
  \hline
\end{tabular}
  \begin{description}
   \item[$^a$]  Error of the dereddened flux ratios. Colons indicate errors larger than 40 per cent.  
  \end{description}
\end{minipage}
\end{table*}

  
\section{Physical Conditions and chemical abundances of \ngc} \label{results}     

The physical conditions of the ionised gas: electron temperature, {\elect}, and density, {\elecd}, have been derived 
from the usual CEL ratios, using the 
{\sc iraf} task {\tt temden} of the package {\sc nebular} \citep{shawdufour95} with the atomic dataset compiled by  
\citet{garciarojasetal05}. 
Electron densities have been derived from {\fni} $\lambda\lambda$5198/5200, {\foii} $\lambda\lambda$3726/3729, 
{\fsii} $\lambda\lambda$6717/6731, {\fcliii} $\lambda\lambda$5518/5538, and {\fariv} $\lambda\lambda$4711/4740  
line ratios. We have also determined {\elecd} from the line ratios of several {\ffeiii} lines 
 following the procedure described by \cite{garciarojasetal06}. 
Electron temperatures have been calculated using several sets of auroral to nebular line intensity ratios: 
{\foii} $\lambda\lambda$7319, 7330/{\foii} $\lambda\lambda$3726, 3729, 
{\fsii} $\lambda\lambda$4069, 4076/{\fsii} $\lambda\lambda$6717, 6731, 
{\foiii} $\lambda$4363/{\foiii} $\lambda\lambda$4959, 5007, {\fsiii} $\lambda$6312/{\fsiii} $\lambda\lambda$9069, 9532 
and {\fariii} $\lambda$5192/{\fariii} $\lambda\lambda$7136, 7751. 
Unfortunately, it was not possible to calculate {\elect}({\fnii}) because the auroral {\fnii} $\lambda$5755 \AA\ line 
was in an incomplete spectral order. 

The procedure for the determination of the physical conditions was the following: an initial {\elect}-value of 8,000 K
was assumed in order to derive a first approximation to the different {\elecd} determinations; then these preliminary 
{\elecd}-values were used to recompute {\elect}, and finally, we iterated until convergence to obtain the final 
adopted values of {\elecd} and {\elect}. {\elect}({\foii}) has been corrected from the contribution to 
{\foii} $\lambda\lambda$7319, 7330 due to recombination following the formulae derived by \citet{liuetal00}. 
This correction only contributes a 2.4\% to the intensity of {\foii} $\lambda\lambda$7319, 7330 lines. 
The physical conditions determined for {\ngc} are shown in Table~\ref{physcond}. 

{\ngc} present a non-uniform density structure. \citet{copettietal00,copettietal07} showed that the {\elecd}({\fsii}) presents a strong spatial
variation, with the density ranging from about 1800 {\cmc} at the brightest eastern-central areas to less than 100 {\cmc} at the outer
parts of the nebula. This density variation was interpreted as steep radial gradient, similar to that found in the Orion Nebula \citep{osterbrockflather59}. 
However, the density estimates listed in Table~\ref{physcond} are rather similar and between 1000 and 2000 {\cmc}.
\cite{copettietal07} obtained {\elecd}({\fsii}) $\approx$ 1200 {\cmc} at the position of the present observations and an upper limit of {\elecd}({\fcliii}) of about 2100 {\cmc} from the 
integrated spectrum. Those authors derive values of {\elect}({\fnii}) and {\elect}({\foiii}) of 11000 and 
9000 K, respectively, consistent with our determinations. 

  \begin{table}
  \begin{minipage}{75mm}
   \centering
   \caption{Physical Conditions of {\ngc}.}
   \label{physcond}
    \begin{tabular}{llc}
     \hline
 & Lines & Value \\  
     \hline
{\elecd} ({\cmc}) & {\fni} &1500$\pm$700 \\
 & {\foii} & 1570$\pm$280 \\
 & {\fsii} & 1140$\pm$230 \\
 & {\ffeiii} & 2330$\pm$800 \\
  & {\fcliii} & 1900$\pm$600 \\
 & {\fariv} & 1670$^{+4900}_{-1670}$ \\
 $T_e$ (K) & {\foii} & 10450$\pm$250 \\
 & {\fsii} & 7840$\pm$360 \\
 & {\foiii} & 9410$\pm$160 \\
 & {\fsiii} & 10760$\pm$230 \\ 
 & {\fariii} & 8650$\pm$280 \\ 
     \hline
    \end{tabular} 
     \end{minipage}
  \end{table}

  Ionic abundances of \ion{N}{+}, \ion{O}{+}, \ion{O}{2+}, \ion{Ne}{2+}, \ion{S}{+}, 
  \ion{S}{2+}, \ion{Cl}{+},  \ion{Cl}{2+},  \ion{Cl}{3+},  \ion{Ar}{2+} and \ion{Ar}{3+} have been derived 
  from CELs under the two-zone scheme and {\tf} $=$ 0, using the {\sc nebular} package, except in the case of \ion{Cl}{+} 
  for which we have used another analysis package (see below). 
  We have assumed {\elect}({\foii}) for the abundance calculations of low ionization potential 
  ions: \ion{N}{+}, \ion{O}{+} and \ion{S}{+}; and {\elect}({\foiii}) for the high 
  ionization potential ones: \ion{O}{2+}, \ion{Ne}{2+}, \ion{S}{2+}, \ion{Cl}{2+}, 
  \ion{Ar}{2+} and \ion{Ar}{3+}. We have adopted {\elecd} = 1360 $\pm$ 170 for all the ions. This value is the weighted mean of the 
  densities obtained from {\fsii}, {\foii} and {\fcliii} line ratios. For \ion{Ne}{2+} we have used the updated atomic data listed in \citet{garciarojasetal09} 
  instead of those included by default in {\sc nebular}.  The older atomic dataset gave an inconsistency of about 0.15 dex in the abundances 
  obtained from the two {\fneiii} lines observed.  Many {\ffeiii} lines have been detected in the 
  spectrum of \ngc. We have calculated the \ion{Fe}{2+}/\ion{H}{+} ratio using 
  a 34-level model atom that includes the collision strengths from \cite{zhang96},  
  transition probabilities of \cite{quinet96} as well as the transitions found by 
  \cite{johanssonetal00}. The average value of \ion{Fe}{2+} abundance has been obtained 
  from 12 individual emission lines and assuming {\elect}({\foii}) as the representative 
  temperature for this ion. We have detected {\ffeii} 4287 \AA, the only line of this ion 
  present in our spectrum. The intensity of this line is affected by continuum or starlight 
  fluorescence \citep[see][]{rodriguez99} and therefore we can not determine an accurate abundance from this 
  line. Unfortunately, 
  the brightest and less sensitive to fluorescence {\ffeii} line, {\ffeii} 
  8616 \AA\ is in one of our observational gaps. Several [{\feiv}] lines have been detected in {\ngc} however only 
  [{\feiv}] 6739.8 \AA\  is not blended with other lines. Using that single line, the Fe$^{3+}$/H$^{+}$ ratio has been derived using a 33-level 
  model atom where all collision strengths are those calculated by \citet{zhangpradhan97} and the transition probabilities recommended 
  by \citet{froesefischeretal08}. We have asummed  {\elect}({\foiii}) to derive the  Fe$^{3+}$ abundances. The \ion{Cl}{+} abundance cannot be derived from the {\sc nebular} routines,
  instead we have used {\sc pyneb} \citep{luridianaetal12}, which is an updated version of the {\sc nebular} package written 
  in {\sc python} programming language. For \ion{Cl}{+} we have used the line wavelengths and energy levels obtained by \citet{radziemskikaufman74}, the transition probabilities compiled 
  by \citet{mendoza83} and the collision strengths from \cite{tayal04}. Ionic abundances are presented in Table~\ref{abundances}.

  \begin{table}
   \begin{minipage}{75mm}
   \centering
     \caption{Ionic and total abundances$^{\rm a}$}
          \label{abundances}
    \begin{tabular}{lccc}
     \hline
        & \multicolumn{2}{c}{This work} & C07$^{\rm b}$ \\
     	 & {\tf} = 0.000 & {\tf} = 0.045 & {\tf} = 0.000 \\
	 & & $\pm$ 0.007 & \\
     \hline
	\multicolumn{4}{c}{Ionic abundances from CELs$^{\rm c}$} \\
     \hline
	N$^+$ & 6.62$\pm$0.02 & 6.75$\pm$0.03 & 6.63$\pm$0.12 \\
	O$^+$ & 7.70$\pm$0.03 & 7.84$\pm$0.03 & 7.65$\pm$0.16 \\
	O$^{2+}$ & 8.19$\pm$0.02 & 8.46$\pm$0.03 & 8.31$\pm$0.06 \\
	Ne$^{2+}$ & 7.23$\pm$0.03 & 7.53$\pm$0.06 & 7.59$\pm$0.08 \\
	S$^+$ & 5.24$\pm$0.02 & 5.36$\pm$0.03 & 5.30$\pm$0.12 \\
	S$^{2+}$ & 6.54$\pm$0.02 & 6.84$\pm$0.06 & 6.78$\pm$0.08 \\
	Cl$^{+}$ & 3.38$\pm$0.04 & 3.49$\pm$0.06 & $-$ \\
	Cl$^{2+}$ & 4.80$\pm$0.02 & 5.06$\pm$0.05 & 5.03$\pm$0.05 \\
	Cl$^{3+}$ & 2.78$\pm$0.07 & 3.01$\pm$0.10 & $-$ \\
	Ar$^{2+}$ & 6.14$\pm$0.02 & 6.37$\pm$0.04 & $-$ \\
	Ar$^{3+}$ & 3.66$\pm$0.09 & 3.94$\pm$0.10 & $-$ \\
	Fe$^{2+}$ & 5.27$\pm$0.10 & 5.40$\pm$0.10 & $-$ \\
	Fe$^{3+}$ & 5.14$\pm$0.29 & 5.37$\pm$0.30 & $-$ \\
     \hline
	\multicolumn{4}{c}{Ionic abundances from RLs$^{\rm d}$} \\
     \hline
	He$^+$ & \multicolumn{2}{c}{10.94$\pm$0.01} & 10.96$\pm$0.04 \\
	C$^{2+}$ & \multicolumn{2}{c}{8.13$\pm$0.05} & $-$ \\
	O$^{+}$ & \multicolumn{2}{c}{7.85$\pm$0.07} & $-$ \\
	O$^{2+}$ & \multicolumn{2}{c}{8.46$\pm$0.03} & $-$ \\
     \hline
	\multicolumn{4}{c}{Total abundances from CELs$^{\rm c}$}\\
     \hline
	N & 7.31$\pm$0.04 & 7.54$\pm$0.05 & 7.37$\pm$0.06 \\
	O & 8.31$\pm$0.02 & 8.55$\pm$0.03 & 8.39$\pm$0.05 \\
	Ne & 7.35$\pm$0.02 & 7.62$\pm$0.07 & 7.68$\pm$0.06 \\
	S & 6.64$\pm$0.02 & 6.93$\pm$0.06 & 6.91$\pm$0.11 \\
	Cl & 4.82$\pm$0.02 & 5.08$\pm$0.05 & 5.16$\pm$0.08 \\
	Ar & 6.17$\pm$0.02 & 6.38$\pm$0.04 & $-$ \\
	Fe & 5.74$\pm$0.08 & 5.94$\pm$0.08 & $-$ \\
     \hline
	\multicolumn{4}{c}{Total abundances from RLs$^{\rm d}$} \\
     \hline
     	He & \multicolumn{2}{c}{10.96$\pm$0.01} & 10.96$\pm$0.04 \\
	C & \multicolumn{2}{c}{8.30$\pm$0.05} & $-$ \\ 
	O & \multicolumn{2}{c}{8.56$\pm$0.03} & $-$ \\ 	   
     \hline
    \end{tabular}
    \begin{description}
      \item[$^a$] In units of 12+log(\ion{X}{+n}/\ion{H}{+}) .  
      \item[$^b$] Copetti et al. (2007).
      \item[$^c$] CELs: Collisionally excited lines.
      \item[$^d$] RLs: Recombination lines.
    \end{description}
   \end{minipage}
  \end{table}

We have measured several {\hei} emission lines in the spectrum of \ngc. 
These lines arise basically from recombination but they can be affected by collisional excitation and 
self-absorption effects. 
We have used the effective recombination coefficients of \citet{porteretal05} -- with the interpolation formulae provided 
by \citet{porteretal07} -- for {\hei} as well as those of \citet{storeyhummer95} 
for {\hi}, in order to calculate the \ion{He}{+} abundance. The collisional contribution
was estimated from \citet{saweyberrington93} and \citet{kingdonferland95}, and the optical depth in 
the triplet lines were derived from the computations by \citet{benjaminetal02}. 
We have determined the He$^+$/H$^+$ ratio from a maximum likelihood method \citep[MLM, ][]{peimbertetal00, 
apeimbertetal02}. 

To determine the He$^+$/H$^+$ ratio, {\elect}(\hei), the temperature fluctuations parameter \citep{peimbert67} for 
{\hei}, \tf(\hei), and the optical depth in the {\hei}  
$\lambda$3889 line, $\tau_{3889}$, in a self-consistent manner, we have used the adopted density obtained from 
the CEL ratios (see Table~\ref{physcond}), and a set of 16 $I$(\hei)/$I$(\hi) line ratios ($\lambda$$\lambda$ 
3614, 3819, 3889, 3965, 4026, 4121, 4388, 4471, 4713, 4922, 5016, 5048, 5876, 6678, 7065, 7281 \AA). We have 
obtained the best sets of values of the three unknowns by minimizing $\chi^2$. The lowest $\chi^2$ parameters 
we obtain are in the range from 14 to 28, which indicate reasonably good fits. The final adopted \ion{He}{+} abundance 
is 12+log(\ion{He}{+}/\ion{H}{+}) = 10.94 $\pm$ 0.01 (see Table~\ref{abundances}). Interestingly, the range of 
values of \tf(\hei) we obtain 
in these fits is between 0.025 and 0.065, consistent with that obtained assuming that the {\adfo} is produced 
by the presence of temperature fluctuations in the ionised gas (see below).

  We have detected several {\cii} lines in our spectrum, but most of them are produced by resonance 
  fluorescence by starlight \citep{estebanetal98,estebanetal04}. Only {\cii} 4267 and 9903 \AA\ are pure RLs and 
  permit to derive the \ion{C}{2+}/\ion{H}{+} ratio. Two RLs of multiplet 1 of {\oi} are detected in the red part of the spectrum, 
  namely {\oi} 7774 and 7775 \AA, and fortunately these important lines are not affected by telluric emission.  
  We also detect and measure a large number of RLs of 
  \oii. In particular, seven lines of multiplet 1, four of multiplet 2 and three of multiplet 10 that 
  are produced by pure recombination and can be used to derive the 
  \ion{O}{2+}/\ion{H}{+} ratio. We have used {\elect}({\foiii}) and the effective recombination coefficients of \cite{daveyetal00} for {\cii} and  
  \cite{storey94} for {\oii} to determine the  \ion{C}{2+} and \ion{O}{2+} abundances; and {\elect}({\foii}) and the effective recombination coefficients of 
  \citet{escalantevictor92} and \citet{pequignotetal91} for the \ion{O}{+} abundance.  
  The \ion{C}{2+} abundance obtained from the {\cii} 4267 and 9903 \AA\ lines are 12 + log (\ion{C}{2+}/\ion{H}{+}) = 8.18 $\pm$ 0.05 
  and 8.10 $\pm$ 0.03, respectively. We adopted the weighted mean of both values for the final \ion{C}{2+} abundance 
  included in Table~\ref{abundances}. In the case of the \ion{O}{+}/\ion{H}{+} we adopted the mean value of the abundances obtained 
  from each of the two lines detected and the two sets of effective recombination coefficients indicated above. 
   Following our usual 
methodology to minimize uncertainties, we have derived the \ion{O}{2+} abundance from the estimated total flux of 
  each multiplet \citep[see][]{estebanetal98}. In particular, we obtain values of 
12+log(\ion{O}{2+}/\ion{H}{+}) = 8.46$\pm$0.03, 8.51$\pm$0.06 and 8.40$\pm$0.07 for multiplets 1, 2 and 10 
respectively, remarkably similar and consistent within the errors. We have assumed the abundance obtained 
from multiplet 1 as the representative one for \ion{O}{2+} (a weighted mean value of the abundances given by each 
multiplet gives almost the same 
value). Our spectrum also shows a relatively large 
number of permitted lines of other heavy-element ions ({\nii}, {\oi} and {\silii}) but they are not 
produced by pure recombination and we can not derive reliable abundances from them 
\citep[see][]{estebanetal98,estebanetal04}.  

  Our deep spectrum permits to calculate \ion{O}{+} and \ion{O}{2+} abundances from two kinds of lines -- RLs and CELs. 
  As for all {\hii} regions where both kinds of lines have been observed \citep[e.g.][]{garciarojasesteban07,estebanetal09}, 
  the abundance determined from RLs is larger 
  than that obtained from CELs.
  Then, we can compute the so-called abundance discrepancy factor for \ion{O}{+}  and \ion{O}{2+}, ADF(X$^{\rm i}$) , which is defined in its 
  logarithmic form as: 
\begin{eqnarray}
 {\rm ADF}({\rm X}^{\rm i}) = {\rm log}({\rm X}^{\rm i}/{\rm H}^+)_{\rm RLs} - {\rm log}({\rm X}^{\rm i}/{\rm H}^+)_{\rm CELs}. 
\end{eqnarray} 
The value of the {\adfo} we obtain for {\ngc} is 0.27 $\pm$ 0.03, remarkably similar 
to the values of this quantity found for other Galactic and extragalactic {\hii} regions, which mean value is 0.26 $\pm$ 0.09
\citep{estebanetal09}. In the case of \ion{O}{+}, we obtain ${\rm ADF(O^{+})}$ = 0.15 $\pm$ 0.08, which is also in excellent agreement 
with the values of between 0.15 and 0.20 we obtain for other well observed Galactic {\hii} regions as M~8, M~20 and the Orion nebula 
\citep{garciarojasesteban07}. 

  If -- as an hypothesis -- we assume the validity of the temperature fluctuations paradigm and that this phenomenon produces 
  the abundance discrepancy \citep[see][]{garciarojasesteban07}, we can estimate the \tf\ parameter 
  that produces the agreement between the abundance of \ion{O}{2+} determined from CELs and RLs,  
  which results to be \tf\ = 0.045 $\pm$ 0.007. 
  In Table~\ref{abundances} we include ionic abundances 
  determined for \tf\ = 0 -- the standard procedure considering no temperature fluctuations -- and 
  assuming \tf\ = 0.045 $\pm$ 0.007. These last calculations have been made following the formalism 
  outlined by \cite{peimbertcostero69}. It is remarkable that the \tf\ parameter value we obtain from the comparison of \ion{O}{+} abundances 
  determined from CELs and RL is 0.048 $\pm$ 0.029, in excellent agreement with the value we obtain for \ion{O}{2+}. This consistency supports that 
  the phenomenon that produce the abundance discrepancy and temperature fluctuations may be connected or even be the same, at least in {\hii} regions 
  \citep[see discussion in][]{garciarojasesteban07}. 

  We have adopted a set of ionization correction factors (ICFs) to correct for the unseen 
  ionization stages and derive the total gas phase abundances of the different elements, except in the cases of O and Cl, for 
  which we have measured emission lines of all the expected ionic species of these elements. 
  The final total abundances for \tf\ = 0 and  \tf\ = 0.045 $\pm$ 0.007 are presented in Table~\ref{abundances}. 
 For He, C, N, S and Ne we have adopted the same ICF schemes used by \citet{garciarojasesteban07} in order to 
 facilitate the comparison with data of other Galactic {\hii} regions and study the abundance gradients. 
 The total helium abundance has been corrected for the presence of neutral helium using the 
  expression proposed by \cite{peimbertetal92} based on the similarity of the ionization potentials 
  of {\ion{He}{0}} and \ion{S}{+}.  
  In the case of C we have adopted the ICF(\ion{C}{+}) 
  derived from photoionization models of \cite{garnettetal99}. In order to derive the total 
  abundance of nitrogen we have used the ICF derived from the models by \citet{mathisrosa91}; note that these ICF values yield 
  N abundances about 0.11 dex higher than those obtained from the usual formulation by \citet{peimbertcostero69} based 
  on the similarity of the ionization potential of \ion{N}{+} and \ion{O}{+}. 
  The total abundance of oxygen is calculated as the sum of \ion{O}{+} and \ion{O}{2+} abundances. 
  The only measurable CELs of Ne in the optical range are those of \ion{Ne}{2+} 
  but the fraction of \ion{Ne}{+} may not be negligible in the nebula. We have adopted the usual expression 
  of \citep{peimbertcostero69} to obtain the total Ne abundance. This scheme seems to be apropriate for the 
  ionization degree of {\ngc}. We have measured CELs of two ionization stages of 
  S: \ion{S}{+} and \ion{S}{2+}, and used the ICF proposed by \cite{stasinska78} to take into account the 
  presence of some \ion{S}{3+}. Chlorine shows lines of the three ionization stages we expect in {\ngc}. Its total abundance is 
  simply the sum of the ionic abundance \ion{Cl}{+},  \ion{Cl}{2+} and  \ion{Cl}{3+}.
  For argon, we have determinations of \ion{Ar}{2+} and \ion{Ar}{3+} but some contribution of 
  \ion{Ar}{+} is also expected. We have adopted the ICF recommended by \cite{izotovetal94} 
  for this element. Finally, we have used an ICF scheme based on photoionization models 
  of \cite{rodriguezrubin05} to obtain the total Fe/H ratio. 
  In Table~\ref{abundances} we include the total abundances 
  determined for \tf\ = 0 and \tf\ = 0.045 $\pm$ 0.007. The variations due to the 
  dependence of the adopted ICFs on the  \tf\ considered are also included  
  in the total abundances given in Table~\ref{abundances}.
  
 In Table~\ref{abundances} we also include the ionic and total abundances obtained by \citet{copettietal07} for  
 their ABC spectrum of {\ngc}. Those authors do not consider temperature fluctuations. We can see that all abundances 
 of twice ionised species determined by \citet{copettietal07} are somewhat larger than our determinations. 
 This fact 
 is perhaps indicating that both sets of observations correspond to nebular areas with somewhat 
 different integrated ionization conditions. However, the total abundances obtained in both studies are in general consistent  
 except in the case of Ne and Cl, where the difference is 0.33 and 0.34 dex, respectively. The discrepancy in Ne/H  
 ratios is due to the large difference between the intensities of the {\fneiii} lines reported in both datasets. In the case of Cl/H 
 the difference could be due to the different methodology used by \citet{copettietal07} and us for determining the abundance of this element. 
 Those authors use \ion{Cl}{2+}/\ion{H}{+} and an ICF and we determine Cl/H directly from the sum of the different ionic abundances.

It is interesting to compare the abundance ratios determined for {\ngc} with those of reference objects. In Table~\ref{ratios} we show the C/O, N/O, Ne/O, 
S/O, Cl/O and Ar/O measured for {\ngc}  in the case of assuming {\tf} = 0 and {\tf} = 0.045, the abundance ratios for the solar photosphere \citep{asplundetal09} -- the protosolar values are identical -- and the values obtained for the Orion nebula \citep[][assuming {\tf} $>$ 0]{garciarojasesteban07}. In general, we can see that the ratios are rather consistent in the 3 objects within the errors. The value we obtain of log(Ne/O) $\sim$ $-$0.93 is somewhat low, even lower than the typical ratio of about $-$0.70 found for extragalactic {\hii} regions covering a very wide range of metalicity \citep{dorsetal13}. This indicates that our determination of 
\ion{Ne}{++}/\ion{H}{+} and Ne/H is probably about 0.20 lower than expected. In the case of the N/O ratio, it is consistent with that obtained by \citet{copettietal07} within the errors, but about 0.17 dex lower than that of the Orion nebula.
 
  \begin{table*}
   \begin{minipage}{155mm}
   \centering
     \caption{Comparison of abundance ratios}
     \label{ratios}
    \begin{tabular}{lcccc}
     \hline
        & \multicolumn{2}{c}{NGC~2579} & &  Orion neb.$^{\rm b}$\\
     	 & {\tf} = 0 & {\tf} $>$ 0 & Sun$^{\rm a}$ & {\tf} $>$ 0 \\
     \hline
	C/O & $-$ & $-$0.26$\pm$0.06 & $-$0.26$\pm$0.07 & $-$0.28$\pm$0.04 \\
	N/O & $-$1.00$\pm$0.04 & $-$1.01$\pm$0.06 & $-$0.86$\pm$0.07 & $-$0.84$\pm$0.10 \\	
	Ne/O & $-$0.96$\pm$0.02 & $-$0.93$\pm$0.08 & $-$0.76$\pm$0.11 & $-$0.77$\pm$0.08 \\	
	S/O & $-$1.67$\pm$0.02 & $-$1.62$\pm$0.07 & $-$1.57$\pm$0.06 & $-$1.48$\pm$0.05 \\
	Cl/O & $-$3.47$\pm$0.02 & $-$3.47$\pm$0.06 & $-$3.19$\pm$0.33 & $-$3.42$\pm$0.05 \\	
	Ar/O & $-$2.14$\pm$0.02 & $-$2.17$\pm$0.05 & $-$2.29$\pm$0.14 & $-$2.08$\pm$0.07 \\		
     \hline
    \end{tabular}
    \begin{description}
      \item[$^a$] Asplund et al. (2009) .  
      \item[$^b$] Garc\'\i a-Rojas \& Esteban (2007).
    \end{description}
   \end{minipage}
  \end{table*}

\section{Chemical Evolution Model and Discussion} \label{discussion}

 The derivation of precise abundances for {\ngc} is an opportunity to explore the behaviour of the radial 
  abundance gradients of ionised gas at galactocentric distances larger than the solar ones. In Figure~\ref{f3} we include the C/H, O/H and C/O 
 ratios of {\ngc} as those for the sample of {\hii} regions of \citet{garciarojasesteban07}, which are also 
 determined  using the intensity of RLs and the same methodology. The objects of that sample are located at galactocentric 
 distances between 6.3 and 10.4 kpc (assuming the Sun at 8 kpc). We also include the radial gradients 
 determined from the linear fit to the data sample of \citet{garciarojasesteban07} but extrapolated up to 13 kpc. As it can be seen in 
 Figure~\ref{f3}, the point of {\ngc} stands above the extrapolated gradients, specially in the 
 case of the C and C/O gradients. Assuming that the distance to this object is well 
 established \citep[that seems to be the case, see discussion of section 5.2 of][]{copettietal07}, our data indicate that the chemical behaviour of 
 {\ngc} does not fit the trend of the inner Galactic {\hii} regions. Although we deal with data of a single object -- and obviously 
 lacking statistical significance -- this result indicates
 that the chemical composition of {\ngc} is consistent with flatten gradients at its galactocentric distance. Another interesting feature of Figure~\ref{f3} is that the C/O gradient seems to be better represented by a step function instead of a linear fit. Regions with R$_G$ $<$ 7.3 kpc show log(C/O) ratios of the ionised gas of about 0.0 but regions farther out from the Galactic Centre can be represented by a constant value 
 around $-$0.26 with a rather small statistical dispersion. 

  \begin{figure}
   \centering
   \includegraphics[scale=0.5]{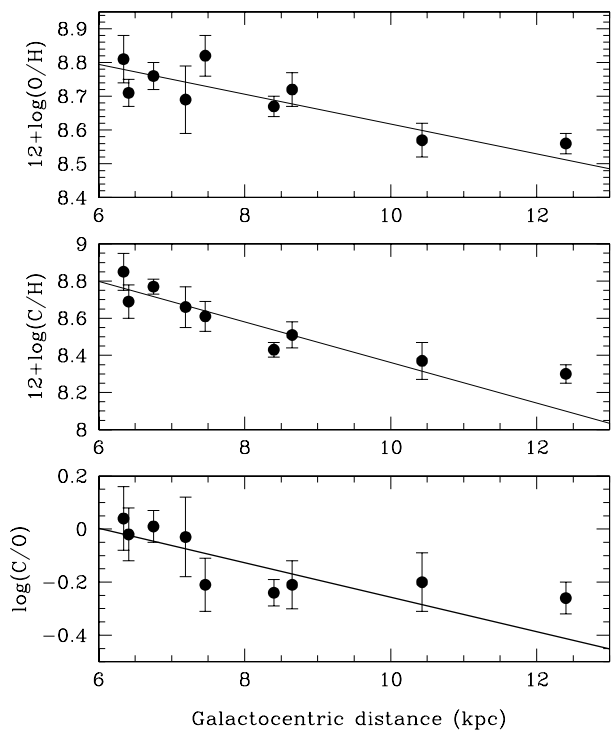} 
   \caption{Ionised gas phase O, C and C/O radial abundance gradients of the Galactic disc from {\hii} region 
   abundances determined from recombination lines. The point at 12.4 kpc corresponds to {\ngc}, the rest are 
   taken from \citet{garciarojasesteban07}. The lines indicate the radial gradients obtained  
   from the linear fit of the data between 6.3 to 10.4 kpc and extrapolated to 13 kpc. The Sun is located at 8 kpc.}
   \label{f3}
  \end{figure}    	 

 Some authors 
 have claimed -- based on both {\hii} regions and planetary nebulae data -- that gas phase radial abundance 
 gradients may flatten out at the outer parts of the Galactic disc 
 \citep{fichsilkey91,vilchezesteban96,macielquireza99,costaetal04,macieletal06}, although other 
 authors do not support such flattening \citep{deharvengetal00,rudolphetal06,henryetal10,balseretal11}. 
 Metallicity determinations of other kinds of objects, such as  Cepheids \citep{lucketal03,andrievskyetal04,yongetal06,pedicellietal09} and  
 Galactic open clusters \citep{twarogetal97,yongetal12}, provide further evidence of such change of slope of the gradients at large distances. 
 Strong evidence for flat radial gradients from the spectra of {\hii} regions in the outer discs of nearby spiral galaxies has been presented in several recent 
 papers by \citet{bresolinetal09} for M83, \citet{goddardetal11} for NGC~4625, \citet{bresolinetal12} for NGC~1512 and NGC~3621 and \citet{werketal11} 
 for a sample of 13 -- mostly interacting -- galaxies. Moreover, a flattening or an upturn of the metallicity gradient has also been found from stellar 
 photometry of red giant branch stars in the outer disc of the spiral galaxies NGC~300 and NGC~7793 \citep{vlajicetal09,vlajicetal11}. 
 The value of 12+log(O/H) -- obtained from the intensity of CELs -- in the outer discs of the above mentioned galaxies ranges between 8.2 and 8.4 \citep{bresolinetal12}, consistent with the  
 value of 8.31 $\pm$ 002 we derive for {\ngc}.  \citet{bresolinetal12} indicate that the flattening of the radial abundance in external spiral galaxies occurs approximately at the 
 isophotal radius, $R_{25}$. In the case of the Milky Way, $R_{25}$  is difficult to estimate. The available determinations give values of 11.5 kpc \citep{devaucouleurspence78} or 13.4 kpc \citep{goodwinetal98}, close to the 
 galactocentric distance determined by \citet{copettietal07} for  {\ngc}. Metallicity studies based on Cepheids and open clusters indicate that the change of the slope of the Fe/H ratio occurs at about 9 kpc  \citep{lepineetal11}. 
 
We have made a first exploration of the possible reasons of the flattening of the abundance gradients in the outer galactic discs building a tailored chemical evolution model. 
\citet{bresolinetal12} discuss several mechanisms that can produce such phenomenon: flattening of the star formation efficiency, radial metal mixing or enriched infall. 
In this paper, we focus our attention on the first possibility because it is the simplest one to implement in our available models. The model we have used is almost identical 
to the intermediate wind yields one (IWY), explained in details by \citet{carigipeimbert11}. That IWY model was built to match three observational
constraints along the Galactc disc: the radial distributions of the surface density of 
the total baryonic mass, $M_{\rm tot}(r )$, and gas mass, $M_{\rm gas}(r )$, and the radial gradient of the O/H ratio 
defined by {\hii} regions of the Milky Way located between 6.3 and 10.4 kpc and derived using RLs.

The main assumptions and ingredients of the IWY model are described in the following. The Milky Way disc was formed in an inside-out scenario from primordial infall with time scales $\tau = r$(kpc) $- 2$ Gyr. The initial mass function is that by \citet{kroupaetal93} in the $0.08 - 80$ $M_\odot$ range. We consider an array of metal dependent yields: (a) for low and intermediate mass stars ($0.8 \leq m/M_\odot \leq 8$), we use the yields by \citet{marigoetal96,marigoetal98} and \citet{portinarietal98}; 
(b) for binary stars ($3 \leq m_{bin}/M_\odot \leq 16$), we adopted the yields by \citet{nomotoetal97} in SNIa formulation by \citet{greggiorenzini83}.
We used $A_{bin} = 0.08$, as the fraction of binary stars that are progenitors of SNIa; (c) for massive stars ($8 \leq m/M_\odot \leq 80$)
we considered the yields including stellar rotation by \citet{hirschi07} and \citet{meynetmaeder02} for $Z \leq 0.004$; and the intermediate wind yields, obtained as an average 
of the yields by \citet[][high mass-loss rate]{maeder92} and \citet[][low mass-loss rate]{hirschietal05}) for $Z = 0.02$. 
Since Fe yields are not computed by the Geneva group, we adopted \citet{woosleyweaver95} values, following \citet{carigihernandez08} prescription.

Our model differs from IWY only in  the assumption of a variable star formation process efficiency, $\nu$, for the outer galactocentric radius. With this 
parameter, the star formation rate, $SFR$, is a spatial and temporal function of the form
\begin{equation}
    {\rm SFR}(r,t) = \nu(r) \times M_{\rm gas}^{1.4}(r,t) \times (M_{\rm gas} + M_{\rm star} )^{0.4}(r, t);
  \end{equation}    
where $M_{\rm star}$ is the surface mass density of stars. The formation process efficiency $\nu$ was kept constant in the IWY model, while in the present paper we assume that $\nu$ depends on the galactocentric radius. This is the only difference between the IWY model and the present one. We adopt $\nu =  0.016 \equiv \nu_{\rm in}$ for $r < 10$~kpc and
$\nu/ \nu_{\rm in}$ = 0.8, 1.3 and 2.2 for  $r$ =10, 12 and 14 kpc, respectively.  The values of $\nu$ as a function of the galactocentric radius we have chosen are those that reproduce the behaviour of O/H radial gradient at large galactocentric radii. 

  \begin{figure*}
   \centering
   \includegraphics[scale=0.70]{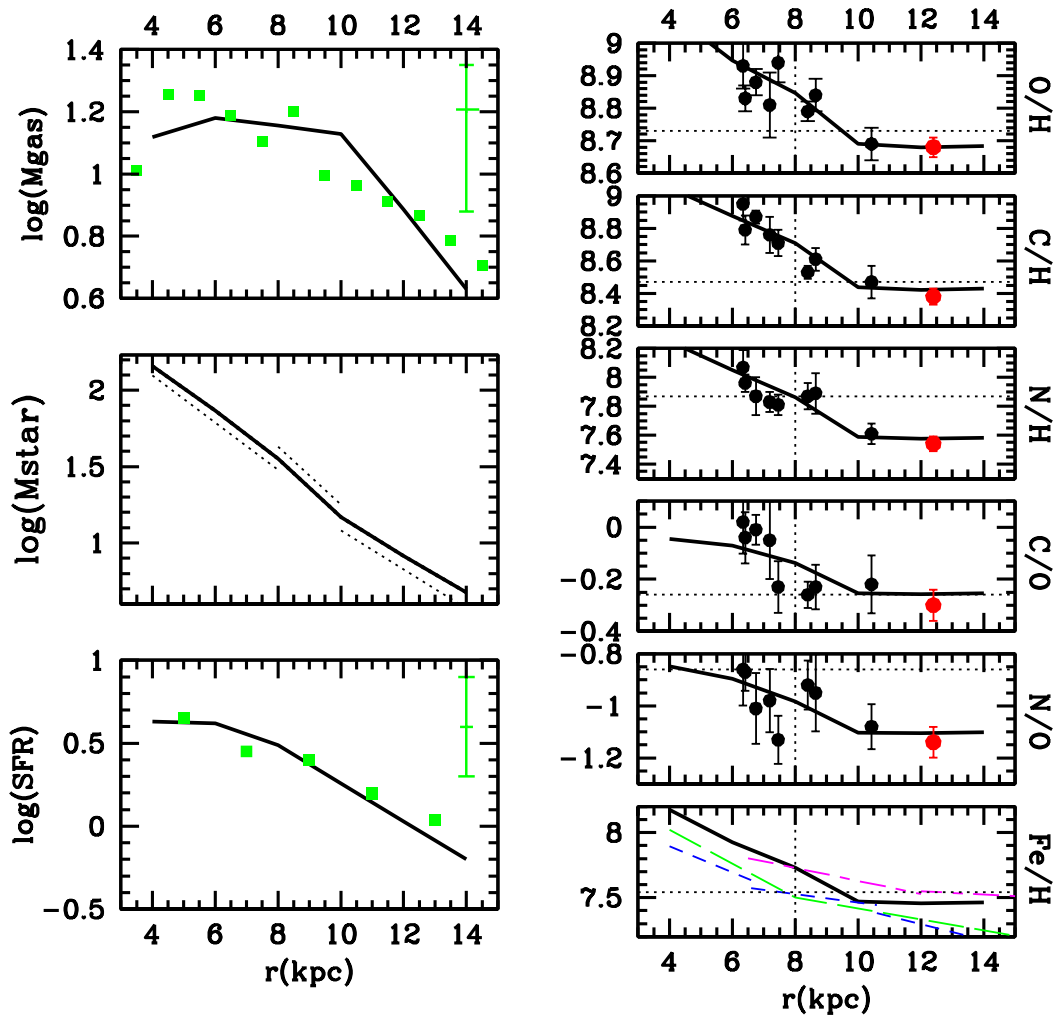} 
   \caption{The present-day radial distribution of surface mass densities of gas ($M_{gas}$) and stars ($M_{star}$) in $M_\odot$ pc$^{-2}$ units and star formation rate ($SFR$)  
in $M_\odot$ pc$^{-2}$  Gyr$^{-1}$ units (left panels); ISM abundance ratios and stellar Fe/H gradient (right panels). 
Continuous black lines: Results of the chemical evolution model that assumes an increase of $\nu$ for $r \geq $10 kpc.
The three dotted lines parallel to the predicted $M_{star}$ correspond to scale lengths ($R_d$)
of 2.8, 2.3 y 3.4 kpc for 4-8 kpc, 8-10 kpc and 10-14 kpc, respectively. Observational data: filled circles represent  {\hii} regions \citep[corrected for dust depletion,][this work]{garciarojasesteban07}; 
red circles: NGC 2579 (corrected for dust depletion, this paper);
discontinuous lines: Fe/H gradient by \citet[][short-long-dashed magenta line]{yongetal12}, \citet[][short-dashed blue line]{lucketal11} and \citet[][long-dashed green line]{pedicellietal09}. 
Filled green squares: values of $M_{gas}$ and $SFR$ compiled by \citet{kennicuttevans12}. Vertical bars: average observational errors. 
Diagrams showing O/H, C/H, N/H and Fe/H express the abundances in units of 12+log(X$_{\rm i}$/H) and those showing C/O and N/O are in 
log(X$_{\rm i}$/O) units. 
}
   \label{cem}
  \end{figure*}

In Figure~\ref{cem}, the model is compared to observational constraints, as follows:
 \begin{itemize}
\item[a)] $M_{gas} (r )$ values are obtained by adding the atomic and molecular data from figure 7 by \citet{kennicuttevans12}. These original data include the hydrogen and helium components.
\item[b)] $SFR(r )$ values are taken from the same figure 7 by \citet{kennicuttevans12}.
\item[c)] The disc scale length, $R_d$, of the Galactic stellar disc ($M_{star}(r) \propto e^{-r/R_d}$) by \citet{yinetal09}. In their table 7 they show $R_d$ values between 2.3 and 5 kpc for $K$, $R$, $V$ and $B$ bands.
\item[d)] O/H and C/H values from {\hii} regions and computed from RLs \citep[][this work]{garciarojasesteban07}. 
The total gaseous abundances shown in our Table~\ref{abundances} have been modified under the assumption that 35\% of the O atoms and 25\% of the C atoms are trapped in dust grains \citep{peimbertpeimbert10}, increasing the O/H and C/H values by 0.12 and 0.10 dex, respectively.
\item[e)] N/H values computed from CELs and assuming {\tf} $>$ 0 \citep[][this work]{garciarojasesteban07}. We have assumed no dust depletion for N.
\item[f)] Fe/H gradients from open clusters and Cepheids by \citet{pedicellietal09}, \citet{lucketal11} and \citet{yongetal12}.
 \end{itemize}  
 
It is remarkable the agreement with the observational constraints for $r \ge 10$ kpc,
specifically with the current C/H, N/H, C/O and N/O abundance gradients (slopes and absolute values) and with the $SFR$, $M_{gas}$ and $M_{star}$ radial distributions, for which the model was not built at all. 

The results of the chemical evolution model we have built in this work indicate that an increase of $\nu$ for $r \ge 10$ kpc can explain the chemical flattening in the outer Galactic disc. It is important to remark that $\nu$ does not represent the star formation efficiency, $SFE$, defined as 
$SFE (r) = SFR (r) / M_{gas}(r)$, the $SFR$ per unit of surface mass density of gas. In order to clarify this point, we show in Figure~\ref{nu} the radial distribution of  $\nu$ and $SFE$.
In this figure, we note that a constant value of  $\nu$ at inner radii implies a decrease of $SFE$ and an increase of $\nu$ at outer radii implies a flat $SFE$. A similar behaviour of the $SFE(r)$ was found by \citet{bigieletal10}, based on the combination of atomic hydrogen and far-ultraviolet emission in 17 spiral and 5 dwarf galaxies. They  conclude that the $SFE$ in the outer disc ($r > R_{25}$) is flatter compared to the $SFE$ in the inner disc ($r < R_{25}$), 
due to the increase of the $SFR$ with {\hi} column density. The $SFE$ is the inverse of the gas depletion time, the time requiered for present-day star formation to consume the available gas. The values of $SFE$ we obtain at the outer disc indicate a depletion time of about 7.5 Gyr and that this value becomes almost constant in that part of the Galaxy. \citet{bigieletal10} find depletion times between 10 to 100 Gyr for their sample of spiral and dwarf galaxies. 

As \citet{bresolinetal12} point out, a flat $SFE$ can approximately translate into a flattening of the abundance gradient, using their equation 1: 
\begin{equation}
    \frac{\rm O}{\rm H} = \frac{(y_0 \times t \times \Sigma_{\rm SFR})}{\mu \times \Sigma_{\rm HI}} \propto SFE; 
  \end{equation}  
where $y_0$ is the net oxygen by mass \citep[0.006, integrated yield obtained from the IWY model, ][]{carigipeimbert11}, $\mu$ = 11.81 corresponds to the conversion factor from abundance by number to abundance by mass; $t$ is the timescale for the star formation and $\Sigma_{SFR}$ and $\Sigma_{HI}$ are the $SFR$ and {\hi} surface density, respectively. Assuming a constant $SFR$ and gas density and that they have been equal to the present-day values in the last Gyrs, we can estimate $t$ from the equation given above. From Figure~\ref{cem}, at the distance of {\ngc}, we obtain  $\Sigma_{SFR} \sim$ 1 $M_\odot$ pc$^{-2}$  Gyr$^{-1}$ and $\Sigma_{HI} \sim \Sigma_{gas} \sim$ 8 $M_\odot$ pc$^{-2}$. With these numbers we obtain that the time required to enrich the ISM up to the oxygen abundance of {\ngc} determined from CELs and adding 0.12 dex due to dust depletion, 12 + log (O/H)$_{\rm CELs}$ = 8.43, is $\sim$ 4.2 Gyr (3.2 Gyr without dust correction). In the case we use the abundance determination based on RLs including dust correction, 12 + log (O/H)$_{\rm RLs}$ = 8.67, the time is $\sim$ 7.5 Gyr (5.7 Gyr without dust correction). Simulations of galaxy formation in a $\Lambda$ cold dark matter ($\Lambda$CDM) universe predict that the outer discs of galaxies could have mean formation times around 4-6 Gyr, while the inner parts should have much longer values, $>$ 10 Gyr \citep{scannapiecoetal08,scannapiecoetal09}. Our estimations for the outer Galactic disc are roughly consistent with these predictions. \citet{bresolinetal12} determine enrichment timescales for the external discs of two spiral galaxies, for the apparently isolated spiral NGC~3621 they obtain a time of $\sim$ 10 Gyr, and a comparatively much shorter value  -- around 2-3 Gyr -- for the interacting spiral galaxy NGC~1512. 

In an upcoming paper, we explore other factors that may produce flat gradients in the outer Galactic disc, as homogeneous infall, enriched accretion, and gaseous and stellar radial migration \citep[see][for references]{bresolinetal12}.

  \begin{figure}
   \centering
   \includegraphics[scale=0.40]{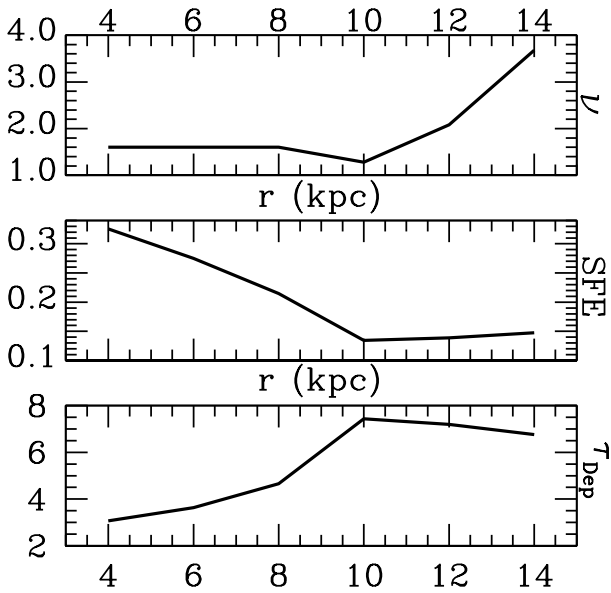} 
   \caption{The present-day radial distribution of:  $\nu$, where $\nu (r) = SFR(r) \times M_{gas}^{-1.4}(r) \times (M_{gas} + M_{star})^{-0.4}(r)$,  in 
0.01 Gyr$^{-1} (M_{\odot}$ pc$^{-2}$)$^{-0.8}$ units (upper panel), the star formation efficiency, $SFE(r) = SFR(r )/M_{gas}(r )$, in Gyr$^{-1}$ units (middle panel) and depletion time $\tau_{\rm Dep}(r) = 1/SFE(r)$ in Gyr (lower panel).}
   \label{nu}
  \end{figure}

\section{Conclusions} \label{conclusions}
 
 We present deep echelle spectrophotometry in the 3550--10400 \AA\ range of the Galactic {\hii} region {\ngc}. This object is located at a rather large galactocentric distance, 12.4 $\pm$ 0.7 kpc 
 \citep{copettietal07} and has been largely neglected due to identification problems finally resolved by \citet{copettietal07}. It also shows a rather high surface brightness and  
 ionization degree. All these properties make {\ngc} an excellent object to explore the behaviour of the Galactic radial abundance gradients of the ionised gas in the outer disc of the Milky Way. 
 
 We have derived consistent and  precise 
 values of the physical conditions of the nebula making use of several emission line-intensity ratios as well as abundances for several ionic species from the intensity of collisionally excited lines (CELs). 
 Our deep spectra permits -- for the first time for a Galactic {\hii} region with so low metallicity -- to obtain a very good determination of the ionic \ion{C}{2+}, \ion{O}{+} and \ion{O}{2+} abundances -- as well as the total O abundance -- from the intensity of faint pure recombination lines (RLs). 
 The comparison between the \ion{O}{+}/H$^{+}$ and O$^{2+}$/H$^{+}$ ratios 
 determined from RLs and CELs gives an abundance discrepancy factor (ADF) of 0.15 $\pm$ 0.08 and 0.27 $\pm$ 0.03 for \ion{O}{+} and \ion{O}{2+}, respectively. Values 
 in complete agreement to the ADFs found for other Galactic and extragalactic 
 {\hii} regions. The value of {\ts} that produces the agreement between the O$^{2+}$ abundance determined from CELs and RLs is {\ts} = 0.045 $\pm$ 0.007, which is consistent to the values we obtain 
from the comparison of the O$^{+}$ abundances  -- {\ts} = 0.048 $\pm$ 0.029 -- and applying a maximum likelihood method for minimising the dispersion of the He$^{+}$/H$^{+}$ ratio from individual lines. 
 
 Our abundance results for {\ngc} based on RLs permit to extend the previous determinations of the C, O and C/O gas phase radial gradients of the inner Galactic disc obtained by \citet{estebanetal05} to 
 larger galactocentric distances.  We find that the chemical composition of {\ngc} is consistent with flatten gradients at its galactocentric distance. This result is in agreement 
 with previous claims of flatten outer abundance gradients based on results for ionised nebulae, Cepheids, and open clusters in the Milky Way and with recent results for the outer discs of nearby spiral galaxies. 
 
We have built a chemical evolution model that reproduces the observed radial distributions of O, C and N abundances, the C/O, N/O ratios and other observational constraints, finding that the flattening of abundance gradients in the Milky Way outside the isophotal radius can be explained due to a shallowed star formation efficiency for $r \ge 10$ kpc.

\section*{Acknowledgments}
We are grateful to the referee, Richard Henry, for comments and suggestions that have improved the paper. 
This work has been funded by the Spanish Ministerio de Educaci\'on y Ciencia (MEC) and Ministerio de Econom\'\i a y Competitividad (MINECO) under projects AYA2007-63030, AYA2010-16717 and AYA2011-22614. LC thanks the funding provided by CONACyT (grant 129753). MVFC thanks the support of the Brazilian agency CAPES. HOC thanks the support of the IPN provided by the project SIP20121451. 

\bibliographystyle{mn2e}
\bibliography{mnrasmnemonic,cesar_bibliography}

\label{lastpage}

\end{document}